\documentclass[a4paper,11pt]{article}
\pdfoutput=1

\usepackage{devanagari}

\def\as(#1){{\alpha_{\rm s}^{\,#1}}}

\usepackage[T1]{fontenc} % if needed

\usepackage{
graphicx,
hyperref,
amsmath,
amssymb,
charter,
xcolor,
ifluatex,
longtable,
booktabs,
multirow,
bbold,cancel}

\usepackage{comment}

\usepackage{tikz}
\usepackage{tikz-feynman}
\tikzfeynmanset{compat=1.1.0}

\usepackage{multirow}

\usepackage{float}
\usepackage{colortbl}

\definecolor{Gray}{gray}{0.95}
\definecolor{RGray}{gray}{0.85}
\definecolor{CGray}{gray}{0.92}

\usepackage{multicol}
\definecolor{tit}{rgb}{0.1,0.2,0.4}
\definecolor{blus}{cmyk}{1,1,0,0.6}
\definecolor{verde}{cmyk}{0.92,0,0.59,0.25}

\usetikzlibrary{matrix}

\usepackage{tipa}
\usepackage{amsmath,amssymb,amsfonts, bm}
\usepackage{dsfont}
\usepackage{epsfig}
\usepackage{graphicx}
\usepackage{slashed}
\usepackage{color}

\usepackage{caption}
\usepackage{subcaption}
\usepackage{slashed} 

\usepackage{adjustbox}
\usepackage{commath}
\usepackage{calc}

\usepackage{sidecap}

\newcommand{\beq}{\begin{equation}}
\newcommand{\eeq}{\end{equation}}

\usepackage{array}

\newcommand{\e}[1]{\cdot 10^{#1}}

\makeatletter
\newcommand*{\rom}[1]{\expandafter\@slowromancap\romannumeral #1@}
\makeatother

\usepackage{float}
\usepackage{caption}

\begin{document}

\allowdisplaybreaks
\vspace*{-2.5cm}
\begin{flushright}
{\small
IIT-BHU
}
\end{flushright}

\vspace{2cm}

\begin{center}
{\LARGE \bf \color{tit} A new determination of   higher-order QCD corrections to hadronic $\tau$ decays  }\\[1cm]

{\large\bf Gauhar Abbas$^{a}$\footnote{email: gauhar.phy@iitbhu.ac.in} 
%M.S.A. Alam Khan$^{b}$\footnote{email:mohdakbar@iisc.ac.in   } \\
Vartika   Singh$^{a}$\footnote{email: vartikasingh.rs.phy19@itbhu.ac.in}    }  
\\[7mm]
{\it $^a$ } {\em Department of Physics, Indian Institute of Technology (BHU), Varanasi 221005, India}\\[3mm]
%{\it $^b$  } {\em }\\[3mm]

\vspace{1cm}
{\large\bf\color{blus} Abstract}
\begin{quote}
We employ \textit{Levin-type sequence transformations} to accelerate the convergence of the perturbative fixed-order expansion of the QCD correction $\delta^{(0)}$ in terms of the strong coupling $\alpha_s$. The method efficiently resums the series, yielding a stable and self-consistent determination of  higher-order QCD corrections to hadronic $\tau$ decays, consistent with existing results. We find $\delta^{(0)}_{\text{Levin-FOPT}} = 0.2089 \pm 0.0040 \, \pm 0.0060_{\alpha_s} \, $, and predict
$
c_{5,1} = 278^{+27}_{-19}, \quad c_{6,1} = 3375^{+489}_{-209}, \quad c_{7,1} = (2.03^{+0.41}_{-0.25}) \times 10^4.
$
Our results demonstrate that Levin-type transformations provide an efficient framework for analyzing asymptotic perturbative series, and  studying  the  higher order perturbative behaviour  of the hadronic $\tau$ decays.

\end{quote}

\thispagestyle{empty}
\end{center}

\begin{quote}
{\large\noindent\color{blus} 
}

\end{quote}

\newpage
\setcounter{footnote}{0}
%\maketitle
%\flushbottom

\def\ca{{C^{}_{\!A}}}
\def\cf{{C^{}_F}}

 \section{Introduction}

The QCD strong coupling constant is one of the most fundamental parameters of the Standard Model (SM) and plays a crucial role in determining the dynamics of strong interactions within and beyond the SM~\cite{Davier:2005xq,Pich:2020gzz}. One of the most accurate low-energy probes of the strong coupling $\alpha_s$ arises from non-strange hadronic $\tau$ decays, which have been extensively investigated over the past decades~\cite{Braaten:1991qm,Baikov:2008jh}. A major stimulus for renewed interest in this channel came from the calculation of the Adler function at four loops~\cite{Baikov:2008jh}. This achievement provided the basis for improved determinations of $\alpha_s(M_\tau)$~\cite{Davier2008}-\cite{Boito:2025pwg}. 

The standard procedure for such analyses is based on the analytic continuation of the Adler function, defined as the logarithmic derivative of the massless QCD polarization function, into the complex energy plane. In this domain, the Adler function can be computed systematically within the operator product expansion (OPE). The impact of higher-dimensional operators in the OPE, the so-called power corrections, on the $\tau$ hadronic width has been evaluated in detail and shown to be numerically suppressed~\cite{Braaten:1991qm,Davier:2005xq,Davier2008,BeJa,Pich2010,Pich_Muenchen}. More recently, attention has shifted toward genuinely nonperturbative contributions that go beyond the OPE description. In particular, possible deviations of the true polarization function from its OPE approximation near the timelike axis, known as quark–hadron duality violations, have been analyzed in a more general framework~\cite{DV}-\cite{Boito:2024gtb}.

Another major source of theoretical uncertainty arises from the renormalization-scale setting in the perturbative expansion. The extraction of the  strong coupling $\alpha_s$ from from non-strange hadronic $\tau$ decays is performed by integrating over  weighted  hadronic spectral functions, which are written as integrals over a closed contour in the complex plane using Finite Energy Sum Rules (FESRs).  This procedure requires a prescription to fix the renormalization-scale.  Two leading frameworks are typically employed for this purpose: fixed-order perturbation theory (FOPT), formulated in terms of $\alpha_s(M_\tau)$, and contour-improved perturbation theory (CIPT), based on contour integrals of $\alpha_s(-s)$~\cite{Pivovarov:1991rh,dLP1}. Applied to the moments of the Adler function, FOPT and CIPT yield systematically different values of $\alpha_s(M_\tau)$~\cite{ParticleDataGroup:2022pth}-\cite{Gracia:2023qdy}, representing the dominant theoretical uncertainty. This discrepancy originates from the asymptotic nature of the QCD perturbative expansion~\cite{Beneke:2008ad}, whose large-order behavior is governed by renormalons~\cite{Beneke:1998ui}. Subtracting the leading infrared renormalon associated with the gluon condensate has been shown to reduce the FOPT–CIPT discrepancy~\cite{Benitez-Rathgeb:2022yqb}-\cite{Beneke:2025hlg}.

Resolving the FOPT-CIPT discrepancy is intrinsically limited by the lack of knowledge of higher-order perturbative contributions. The uncertainty due to the unknown size of these corrections~\cite{Pich:2020gzz,Wu:2019mky} has been addressed through various resummation techniques, including Borel summation, renormalon-based models, conformal mappings, Euler-type transformations, and RG-summed expansions~\cite{Beneke:1998ui}-\cite{Wilson:1969zs}. Since additional higher-order coefficients are unlikely to become available in the near future, analyses of convergence properties and the FOPT-CIPT mismatch within such frameworks are essential for reliable extractions of $\alpha_s$ from inclusive $\tau$ decays.

In this work, we investigate, for the first time, the  unknown higher-order corrections to hadronic $\tau$ decays in the framework of FOPT\footnote{We work with FOPT due to its consistency with the standard OPE and the conventional treatment of nonperturbative effects, as discussed in Refs.~\cite{Hoang:2020mkw,Hoang:2021nlz,Golterman:2023oml,Caprini:2023tfa,Beneke:2025hlg}.} using the Levin-type sequence transformations~\cite{levin:1973}. The Levin transform is a nonlinear sequence transformation designed to accelerate the convergence of slowly convergent series and to enable the summation of divergent ones. Its key feature is the explicit use of remainder estimates, which incorporate information about the asymptotic behavior of the series terms into the transformation. By exploiting this additional structure, the Levin transform is particularly effective for series with factorial or logarithmic divergence, where conventional partial sums converge poorly~\cite{weginer:1989}.  The Levin-type sequence transformations are also employed in perturbation theory  in Refs. \cite{Jentschura:1999vn,Jentschura:2000iw,David:2013gaa}

This paper is organized as follows. In Sec.~\ref{sec1}, we discuss the hadronic decay of the $\tau$ lepton and introduce the FOPT and CIPT expansions. The Levin transformation and its main characteristics are described in Sec.~\ref{levin_trans}. In Sec.~\ref{higher_ord}, we present the investigation of higher-order QCD corrections to hadronic $\tau$ decays. The final determination of higher-order corrections to hadronic $\tau$ decays  is given in Sec.~\ref{higher_orders}. Our estimate of the true value of $\delta^{(0)}$ is discussed in section \ref{delta0}. A summary and concluding remarks are presented in Sec.~\ref{sum}.

\section{The hadronic decay of the $\tau$ lepton}
\label{sec1}

The hadronic decay rate of the $\tau$ lepton provides one of the most precise deterimination of the strong coupling constant $\alpha_s$.  The hadronic decay width of the $\tau$ lepton can be decomposed experimentally into three components: a vector ($V$), an axial-vector ($A$), and a strange contribution ($S$), corresponding respectively to final states produced by the non-strange $\bar u d$ current and the strange $\bar u s$ current. These  partial hadronic widths normalized to the purely leptonic decay width $\Gamma(\tau \to \nu_\tau \, e^{-} \,\bar\nu_e)$, define the dimensionless ratios $R_{\tau,V}$, $R_{\tau,A}$ and $R_{\tau,S}$. The extraction of $\alpha_s$ is most reliable from  the non-strange component of the hadronic tau decay width due to the negligible  corrections from  the masses of the light quarks. Therefore,  we restrict our analysis to the $V$ and $A$ channels.

The inclusive non-strange decay rate can be written as:
\begin{equation}
R_{\tau;V/A}
=
\frac{\Gamma(\tau \to \nu_\tau + \mathrm{hadrons}_{V/A})}
     {\Gamma(\tau \to\nu_\tau \, e^{-} \,\bar\nu_e)}
=
\frac{N_C}{2}\, |V_{ud}|^2\, S_{EW}
\left[
1 
+ \delta^{(0)}
+ \delta_{EW}^{\prime}
+ \delta^{(2,m_q)}_{ud,V/A}
+ \sum_{D \geq 4} \delta^{(D)}
\right],
\label{Eq:Rtau}
\end{equation}
where $S_{EW} = 1.01907 \pm 0.0003$~\cite{ParticleDataGroup:2022pth}  denotes  electroweak corrections~\cite{Marciano:1988vm,Davier:2002dy}, while $\delta_{EW}^{\prime} = 0.0010$~\cite{Braaten:1990ef} represents residual non-logarithmic electroweak corrections. $\delta^{(0)}$ is the perturbative QCD correction, which provides the dominant contribution ($\sim 20\%$). The dimension-$D=2$ quark-mass term $\delta_{ud,V/A}^{(2,m_q)}$ is negligible ($<0.1\%$ for $u,d$), whereas the higher-dimensional $\delta^{(D)}$ encodes OPE condensate contributions and potential duality-violating effects~\cite{ALEPH:2005qgp,Davier:2013sfa,Pich:2022tca,Boito:2024gtb}.

To analyse the hadronic decay width of the $\tau$ lepton, it is convenient to introduce the vacuum two–point correlation functions of the charged weak currents. They are defined as,
\begin{equation}
\label{PiVAmunu}
\Pi_{\mu\nu,ij}^{V/A}(p)
=
i\!\int d^4x\,e^{ipx}\,
\langle \Omega |\,T\left\{J_{\mu,ij}^{V/A}(x)\,J_{\nu,ij}^{V/A}(0)^\dagger\right\}| \Omega \rangle ,
\end{equation}
where $| \Omega \rangle$ represents the physical vacuum,  $J_{\mu,ij}^{V/A}(x)=\bar q_j\gamma_\mu(\gamma_5)q_i(x)$ denote the vector or axial–vector currents constructed from light quark fields with flavour indices $i,j=u,d$. Lorentz covariance implies that the correlators can be decomposed into angular momentum $J=1$(transverse) and angular momentum $J=0$ (longitudinal) components in the hadronic rest frame,
\begin{equation}
\Pi_{\mu\nu,ij}^{V/A}(p)
=
(p_\mu p_\nu - g_{\mu\nu}p^2)\,\Pi^{V/A,(1)}_{ij}(p^2)
+
p_\mu p_\nu\,\Pi^{V/A,(0)}_{ij}(p^2),
\end{equation}
where $\Pi^{V/A,(1)}$ and $\Pi^{V/A,(0)}$ correspond, respectively, to the transverse and longitudinal hadronic contributions.

By unitarity, the inclusive hadronic decay rate can be expressed as a weighted integral of the spectral function of $\Pi^{V/A(1+0)}(s)$ along the timelike axis. Using analyticity and Cauchy’s theorem \cite{Braaten:1991qm},$R_{\tau;V/A}$  may be rewritten as a contour integral in the complex $s$-plane, conveniently chosen as the circle $|s|=M_\tau^2$. After integration by parts one finds the perturbative contribution to $R_{\tau;V/A}$ as,  
\begin{equation}
\label{del0}
\delta^{(0)} = \frac{1}{2\pi i} 
\oint\limits_{|s|=M_\tau^2} \frac{ds}{s}\,
\left(1 - \frac{s}{M_\tau^2}\right)^{\!3} 
\left(1 + \frac{s}{M_\tau^2}\right) 
\widehat{D}_{\rm pert}(a,L).
\end{equation}
The reduced Adler function $\widehat{D}(s) \equiv D^{(1+0)}(s) - 1 $ is derived  by  the logarithmic derivative of the polarization function, 
\begin{align}
D^{(1+0)}(s) & \equiv -\,s\,\frac{d\Pi^{(1+0)}(s)}{ds}, 
\end{align}
where the superscript stands for the spin~\cite{Braaten:1991qm}.

The reduced function $\widehat{D}(s)$  can be expanded as,
\begin{equation}
\label{Ds}
\widehat{D}_{\rm pert}(a,L) 
= \sum_{n=1}^\infty a^n 
\sum_{k=1}^n k\,c_{n,k}\,L^{\,k-1},
\end{equation}
where
\begin{equation}
\label{aL}
a \equiv \frac{\alpha_s(\mu^2)}{\pi}, 
\qquad 
L \equiv \ln \left(-\frac{s}{\mu^2}\right).
\end{equation}

The coefficients $c_{n,1}$ are the independent coefficients, which require $(n+1)$-loop calculations, and the coefficients with $k\geq 2$ follow from the renormalization group, while $c_{n,0}$ encode external renormalization and are not observable. Moreover, $c_{n,n+1}=0$ for $n \geq 1$. The coefficients $c_{0,0} = c_{1,0}=1$.  The nontrivial coefficients $c_{2,1}$, $c_{3,1}$, and $c_{4,1}$ were determined in \cite{Bardeen:1978yd,Baikov:2008jh}, and  for $n_f=3$,
\begin{align}
c_{2,1} &= \tfrac{299}{24} - 9\zeta_{3} = 1.63982 \, , \\ \nonumber 
c_{3,1} &= \tfrac{58057}{288} - \tfrac{779}{4}\zeta_{3} + \tfrac{75}{2}\zeta_{5} = 6.37101 \, , \\ \nonumber 
c_{4,1} &= 49.076 \, .
\end{align}

The QCD $\beta$-function coefficients $\beta_i$ are taken from \cite{Baikov:2008jh,Baikov:2016tgj}; for $n_f=3$,
\begin{align} 
\beta_{0} &= 2.75 - 0.166667 n_{f} = 2.25 , 
\nonumber \\[0.02in]
\beta_{1} &= 6.375 - 0.791667 n_{f} = 4 , 
\nonumber \\[0.02in]
\beta_{2} &= 22.3203 - 4.36892 n_{f} + 0.0940394 n_{f}^{2} = 10.059896 , 
\label{Eq:betai} 
\\[0.02in]
\beta_{3} &= 114.23 - 27.1339 n_{f} + 1.58238 n_{f}^{2} + 0.0058567 n_{f}^{3} = 47.228040 , 
\nonumber \\[0.02in]
\beta_{4} &= 524.56 - 181.8 n_{f} + 17.16 n_{f}^{2} - 0.22586 n_{f}^{3} - 0.0017993 n_{f}^{4} = 127.322 \, . 
\nonumber
\end{align} 
For   $i > 4$, we set  $\beta_i=0$ as  used in previous  studies on higher order expansions \cite{Beneke:2008ad}.

The FOPT expansion of the Adler function is obtained by the choice $\mu^2 =M_{\tau}^2$, and reads as,
\begin{equation}
\label{DsF}
\widehat{D}_{\rm FOPT}(s) 
= \sum_{n=1}^\infty a^n 
\sum_{k=1}^n k\,c_{n,k}\,\left(\ln \dfrac{-s}{M_{\tau}^2} \right)^{\,k-1},
\end{equation}

On the other hand, the CIPT \cite{dLP1, Pivovarov:1991rh} employs the RG-improved expansion obtained by setting $\mu^2=-s$. In this case, Eq.~(\ref{Ds}) simplifies to  
\begin{equation}\label{DsCI}
 \widehat{D}_{\rm CIPT} \left(\tfrac{\alpha_s(-s)}{\pi},0\right) 
   = \sum_{n=1}^\infty c_{n,1}\,\left(\tfrac{\alpha_s(-s)}{\pi}\right)^n .
\end{equation}
% The essential improvement arises from the consistent use of the running coupling $\alpha_s(-s)$, evaluated by solving the renormalization-group equation numerically along the contour, with the initial condition $\alpha_s(M_\tau^2)$ at $s=-M_\tau^2$.  

We take as input the strong coupling at the $Z$-boson mass,
$\alpha_s(M_Z)=0.11873(56)$ ~\cite{Brida:2025gii} .
The strong coupling is evolved to the $\tau$-lepton mass scale using the
\texttt{RunDec} package~\cite{Chetyrkin:2000yt},\cite{Herren:2017osy}, employing renormalization-group evolution
in the $\overline{\mathrm{MS}}$ scheme with heavy-quark threshold matching to
consistently decouple the number of active quark flavours at the bottom- and
charm-quark masses.
Using four-loop running and $\overline{\mathrm{MS}}$ decoupling, we obtain
\begin{equation}
\alpha_s(m_\tau)=0.31959(445)\,.
\end{equation}

The $\delta^{(0)} $ is calculated with the help of the  integrals having the form,
\[
I(q,k) = \frac{1}{2\pi i} \oint_{|s|=s_0} s^q 
\Big(\ln\frac{-s}{\mu^2}\Big)^k ds ,
\]
which evaluate to \cite{Beneke:2008ad,Penarrocha:2001ig}
\[
I(q,k) = s_0^{q+1} \sum_{p = 0}^k 
\sum_{l = 0}^{k-p}  \frac{1 - (-1)^{p}}{2} \,
(-1)^{\tfrac{p-1}{2}} \,
\frac{k!}{p! \, l!} \,
\frac{(-1)^{k-p-l}}{(q+1)^{\,k-p-l+1}} \pi^{p-1} 
\Big(\ln\frac{s_0}{\mu^2}\Big)^{l}, 
\quad q \neq -1,
\]
and
\[
I(-1,k) = 
\sum_{p=0}^{k}\frac{1+(-1)^{p}}{2} 
(-1)^{p/2}\frac{\pi^{p}k!}{(k-p)!\,(p+1)!} 
\Big(\ln\frac{s_0}{\mu^{2}}\Big)^{k-p}.
\]

Setting $\mu^{2}=s_{0}=M_\tau^2$, one obtains the FOPT expansion
\begin{align}
\delta^{(0)}
&= a + 5.20232 \, a^2 + 26.3659 \, a^3 
  + (78.0029 + c_{4,1}) \, a^4 + (-391.546+14.25 c_{4,1}+ c_{5,1}) \, a^5 
\nonumber \\
&\quad + (-7860.51+45.1119\, c_{4,1}+17.8125\, c_{5,1}+c_{6,1}) \, a^6 \, ,
\label{Eq:delta_FOPT} 
\end{align}
where $a=\alpha_s(M_\tau)/\pi$.

In this work, we propose an alternative resummation method based on \textit{Levin-type sequence transformations}, which achieve a substantial acceleration of convergence. This method allows a systematic study of the higher-order structure of the FOPT expansion of $\delta^{(0)}$ in Eq.~\eqref{Eq:delta_FOPT}.

\section{The Levin transform}
\label{levin_trans}
The Levin transform  is a powerful nonlinear sequence transformation capable of  accelerating the convergence of slowly convergent or even strongly divergent series~\cite{levin:1973}. Its central idea is to incorporate explicit estimates of the truncation error into the transformation. It is constructed in such a way that it represents an exact value of   model sequences,
\begin{align}
\label{seq_n}
s_n  =&  s  +  \omega_n \sum_{j=0}^{k-1}  \dfrac{c_j}{ (n +
\beta)^j } , \qquad  n \in \text{N}_0,
\end{align}
where $\text{N}_0 = {0,1,2,3, \cdots}$ and   $\beta$ is an arbitrary parameter.  Here $s$ denotes the full sum i.e. the value which the sequence $s_n$ tends to and which the Levin transform seeks to reconstruct, and $\omega_n$ denotes  the remainder estimate, which is an arbitrary functions of $n$, and depending on its behaviour, the sequesnce $s_n$ may converge or diverge.  We notice that in Eq.\eqref{seq_n}, $\beta+ n$ cannot be zero which requires  $\beta>0$.  For a review of the Levin transform, see Ref. \cite{weginer:1989}.

As demonstrated and emphasized by Smith and Ford in  extensive numerical studies of several linear and nonlinear series transformations,  Levin’s transformations are probably the most effective and versatile convergence accelerators currently available, with the additional capability of summing even strongly divergent series \cite{smith:1979,smith:1982}.  A general representation of the Levin transform is given by \cite{weginer:1989},
\begin{equation}
\label{levin_gen}
\mathcal{L}_k^{(n)}(\beta, s_m, \omega_m) = 
\frac{\displaystyle \sum_{j=0}^k (-1)^j \binom{k}{j} ~ \gamma ~\frac{s_{n+j}}{\omega_{n+j}}}
{\displaystyle \sum_{j=0}^k (-1)^j \binom{k}{j} ~\gamma ~\frac{1}{\omega_{n+j}}},
\end{equation}
where the multiplicative factor is,
\begin{align}
\gamma & = \dfrac{(n+\beta+j)^{k-1}}{(n+\beta+k)^{k-1}}.
\end{align}

In addition to the scaled Levin transform of Eq.~\eqref{levin_gen}, we also employ an unscaled version, which we refer to as the \emph{simplified Levin-transformation}. In Weniger's formulation~\cite{weginer:1989}, the multiplicative factor $\gamma$ is introduced to reduce the magnitude of the numerator and denominator sums and to avoid numerical overflow for larger $k$. The simplified Levin-transformation is obtained by omitting the multiplicative factor $\gamma$ in Eq.~\eqref{levin_gen}, and is given as,
\begin{equation}
\label{levin_simp}
\tilde{\mathcal{L}}_k(s_m,\omega_m)=
\frac{\displaystyle \sum_{j=0}^k (-1)^j \binom{k}{j}\,\frac{s_{j}}{\omega_{j}}}
     {\displaystyle \sum_{j=0}^k (-1)^j \binom{k}{j}\,\frac{1}{\omega_{j}}}.
\end{equation}
The simplified version is useful for low-order checks, while the scaled form provides better numerical stability at higher $k$. Using both forms provides a useful robustness test of the higher-order predictions obtained from the Levin method.

The remainder $ \omega_n$  is chosen in such a way that  $ \omega_n$ is proportional to the dominant term of an asymptotic expansion  $s_n-s$ \cite{weginer:1989}. For a sequence of partial sums,
\begin{align}
   s_n= \sum_{\nu=0}^{n} a_\nu,
\end{align}
there are following standard estimates of the remainder $ \omega_n$ leading to different variations of the Levin transform transform \cite{weginer:1989},
\begin{itemize}
    \item Levin suggested $\omega_n = (\beta + n) a_n$, which provides Levin's $\mathrm{U}$ transform.
    \item  The choice $\omega_n =  a_n$ results in Levin's $\mathrm{T}$ transform.
\item A remarkable choice for strictly alternating terms $a_\nu$ is provided by Smith and Ford \cite{smith:1979} as $\omega_n =  a_{n+1}$ leading to a modified Levin's $\mathrm{T}$ transform which is named as Levin's $\mathrm{D}$ transform.
    
\end{itemize}

We notice that for a sequence of $N$ available partial sums $s_0, s_1, \cdots, s_{N-1}$, the Levin transformation parameters $n$ and $k$ must satisfy
\begin{equation}
n + k \leq N - 1
\end{equation}
to ensure that a sufficient number of input terms are available for constructing the transform. Equality in the above relation corresponds to the case where all available partial sums are utilized in forming the transformation. For the Levin–D variant, where the remainder is defined as $\omega_n = a_{n+1}$, the effective number of usable partial sums is reduced by one, leading to the stricter condition
\begin{equation}
n + k \leq N - 2,
\end{equation}
with equality again indicating the full use of all available terms.

\section{Higher order behavior}
\label{higher_ord}
As discussed earlier, FOPT and CIPT correspond to two distinct prescriptions for renormalization-scale setting. This difference leads to in-equivalent perturbative expansions and, consequently, to different extracted values of $\alpha_s$. The resolution of this ambiguity ultimately depends on the knowledge of the yet uncalculated higher-order corrections. In this section, we estimate these higher-order contributions using Levin-type sequence transformations and compare our results with the existing estimates available in the literature.

%\subsection{Levin transform applied to the $\delta^{(0)}$}

In this section, the higher-order coefficients $c_{5,1}$–$c_{12,1}$ are estimated when  the Levin–U, Levin–T, and Levin–D transformations are applied to the FOPT series for $\delta^{(0)}$ in Eq.~\eqref{Eq:delta_FOPT}. The resulting   Levin–transformed resummed expressions take the analytic form of rational functions, i.e.\ generalized Padé approximants. By re-expanding these Padé representations as a power series in the expansion variable i.e. $a$, one can systematically extract the higher-order coefficients $c_{k,1}$ implied by the corresponding resummation. This provides a self-consistent determination of the unknown  coefficients $c_{5,1}$–$c_{12,1}$ solely from the lower-order input of the original series, without requiring any external information. Different values of the parameter $\beta$ are used within each Levin transformation to examine the stability of the predictions and their sensitivity to this parameter.

To evaluate the predictive performance of Levin transformations, the method is first tested by inputting only three $ (c_{1,1}- c_{3,1})$ of the four exactly known coefficients and predicting the fourth-order coefficient $c_{4,1}$. The percentage deviation of the predicted $c_{4,1}$ from its exact value serves as an indicator of the intrinsic accuracy of each transformation. The higher-order coefficients $c_{5,1}$–$c_{12,1}$ are then predicted using the same three known coefficients as input.  This procedure is employed to identify the most reliable Levin-type transformation, i.e., the one yielding the smallest deviation when only three known coefficients $ (c_{1,1}- c_{3,1})$ are provided as input.  This analysis is followed by a recalculation using four known coefficients to examine the stability and convergence behavior of the method.

The transformations are denoted collectively as $\mathcal{L}_{\mathrm{X},k}^{(n)}(\beta)$, with $\mathrm{X} = \mathrm{T}, \mathrm{U}, \mathrm{D}$ representing the Levin–T, Levin–U, and Levin–D variants, respectively. The simplified form of each transformation, $\tilde{\mathcal{L}}_{\mathrm{X},k}$, corresponds to the case in which the multiplicative factor $\gamma$  is omitted as shown in \ref{levin_simp}, making the transformation independent of that factor and therefore independent of the parameters $\beta$ and $n$. The standard value of the parameter $\beta$ of the Levin-type transform used in literature is $\beta=1$  \cite{weginer:1989}.  However, to test the robustness of our predictions, we use different values of the parameter $\beta$  to study the variation of the predicted coefficients and their sensitivity to $\beta$. Certain Levin transformations are independent of the choice of $\beta$. In such cases the $\beta$ argument plays no role and is therefore omitted from the notation. Consequently, only one numerical entry is shown in the tables for these transforms, as the results corresponding to different values of $\beta$  will coincide and listing them all would be redundant.
 For calculation of the  Levin-sum in tables \ref{tab:coeff_LevinU_1_delta0}, \ref{tab:coeff_LevinT_1_delta0}, and \ref{tab:coeffs_LevinU_2_delta0}-\ref{tab:coeffs_LevinD_2_delta0}, we use  $\alpha_s = 0.31959$.

As discussed earlier, the remainder function of the Levin–U transformation depends explicitly on both $\beta$ and $n$, therefore, no simplified version of the Levin–U transform exists that is independent of these parameters. In contrast, simplified variants independent of  $\gamma$ are presented for the Levin–T and Levin–D transforms.

In Table~\ref{tab:coeff4_LevinU_delta0}, we present the predictions of the fourth-order coefficient $c_{4,1}$ obtained from the Levin–U transformations using only the first three exactly known coefficients, $c_{1,1}$–$c_{3,1}$, as input. Among the various transformations, $\mathcal{L}_{\mathrm{U},2}^{(0)}$ yields the most accurate estimates, with deviations of approximately $6.13\%$ from the exact value. Also, the $n=0,k=2$ Levin-U transform is independent of any value of $\beta$. In contrast, the transformation $\mathcal{L}_{\mathrm{U},1}^{(1)}(1)$ and $\mathcal{L}_{\mathrm{U},1}^{(1)}(5)$ exhibit a significantly larger deviation of about $149.48\%$ and $58.72\%$ respectively, indicating a less reliable performance, hence it is excluded from the final set of predictions. Table~\ref{tab:coeff_LevinU_1_delta0} summarizes the corresponding predictions for the higher-order coefficients $c_{5,1}$–$c_{12,1}$, obtained using the same three input coefficients.

\begin{table}[H]
\centering
\renewcommand{\arraystretch}{1.4}
\setlength{\tabcolsep}{10pt}
\begin{tabular}{c|ccc}
\hline
Coefficient &  $\mathcal{L}_{\mathrm{U},2}^{(0)}/\tilde{\mathcal{L}}_{\mathrm{T},2}$ &  $\mathcal{L}_{\mathrm{U},1}^{(1)}(1)$  & $\mathcal{L}_{\mathrm{U},1}^{(1)}(5)$ \\
\hline\hline
$c_{4,1}$  & $52.08$  & $122.44$   & $77.89$\\
$\Delta c_{4,1}$  & $6.13\%$ & $149.48\%$    & $58.72\%$\\
\hline
\end{tabular}
\caption{Predicted fourth-order coefficient $c_{4,1}$ obtained from the FOPT expansion of $\delta^{(0)}$ using the Levin–U transformation, $\mathcal{L}_{\mathrm{U},k}^{(n)}(\beta)$. The percentage deviation from the exact value is also shown. Since $\mathcal{L}_{\mathrm{U},2}^{(0)}$ coincides with the simplified Levin–T transform $\tilde{\mathcal{L}}_{\mathrm{T},2}$, the corresponding simplified $\mathrm T$-transform result is not listed separately in the Levin–T table.}
\label{tab:coeff4_LevinU_delta0}
\end{table}

\begin{table}[H]
\centering
\renewcommand{\arraystretch}{1.4}
\setlength{\tabcolsep}{10pt}
\begin{tabular}{c|ccc}
\hline
Coefficient  & $\mathcal{L}_{\mathrm{U},2}^{(0)}/\tilde{\mathcal{L}}_{\mathrm{T},2}$  & $\mathcal{L}_{\mathrm{U},1}^{(1)}(1)$  & $\mathcal{L}_{\mathrm{U},1}^{(1)}(5)$ \\
\hline\hline
$c_{5,1}$   & $272.78$ & $170.61$ &$203.35$ \\
$c_{6,1}$   & $3541.38$ & $1.09\e{4}$ & $6174.75$\\
$c_{7,1}$    & $1.94\e4$ & $-2223.32$ &$6475.63$ \\
$c_{8,1}$    & $3.83\e5$ & $1.53\e6$ &$8.00\e5$ \\
$c_{9,1}$   &  $1.43\e6$  & $-6.67\e6$ &$-2.53\e6$\\
$c_{10,1}$   & $7.01\e7$ & $3.66\e8$ & $1.82\e8$\\
$c_{11,1}$  & $-1.38\e8$  & $-4.21\e9$ & $-1.92\e9$\\
$c_{12,1}$  & $2.24\e{10}$ & $1.48\e{11}$ &$7.17\e{10}$ \\
Levin-Sum   & $0.2096$& $0.2780$  & $0.2252$ \\
\hline
\end{tabular}
\caption{Predicted higher-order coefficients $c_{5,1}$–$c_{12,1}$ obtained from the FOPT expansion of $\delta^{(0)}$, estimated using the Levin-U sequence transformation, $\mathcal{L}_{\mathrm{U},k}^{(n)}(\beta)$ using three known coefficients as input. Since $\mathcal{L}_{\mathrm{U},2}^{(0)}$ coincides with the simplified Levin–T transform $\tilde{\mathcal{L}}_{\mathrm{T},2}$, the corresponding simplified $\mathrm T$-transform result is not listed separately in the Levin–T table.}
\label{tab:coeff_LevinU_1_delta0}
\end{table}

In Table~\ref{tab:coeff4_LevinT_delta0} , we present the predictions of the fourth-order coefficient $c_{4,1}$ obtained using the Levin–T  transformation, with only the first three exactly known coefficients $c_{1,1}$–$c_{3,1}$ as input. Among the various Levin-T transformations, the simplified Levin-T transform $\tilde{\mathcal{L}}_{\mathrm{T},2}$ provides the most accurate estimate, with a deviation of $6.13\%$ from the exact value. Since this coincides with the result obtained from the Levin-U transformation $\mathcal{L}_{\mathrm{U},2}^{(0)}$, we do not list these values separately.
 The corresponding predictions for the higher-order coefficients $c_{5,1}$–$c_{12,1}$, obtained using the same three input coefficients, are summarized in Table~\ref{tab:coeff_LevinT_1_delta0}.The $n=1,k=1$ Levin-T transformation does not depend on value of $\beta$.The $n=0,k=1$ Levin-D transform is independent of the choice of value $\beta$ and yields the same result as the $n=1,k=1$ Levin-T transform. Moreover, the simplified Levin-D transform also produces an identical result to the $n=1,k=1$ Levin-T case. Since these values are already reported in the table for the Levin-T transformation, presenting them again for the Levin-D transformation would be redundant, and they are therefore not listed separately.

  \begin{table}[H]
\centering
\renewcommand{\arraystretch}{1.4}
\setlength{\tabcolsep}{10pt}
\begin{tabular}{c|ccc}
\hline
Coefficient & $\mathcal{L}_{\mathrm{T},2}^{(0)}(1)$ & $\mathcal{L}_{\mathrm{T},1}^{(1)}/\tilde{\mathcal{L}}_{\mathrm{D},1}/\mathcal{L}_{\mathrm{D},1}^{(0)}$& $\mathcal{L}_{\mathrm{T},2}^{(0)}(5)$ \\
\hline\hline
$c_{4,1}$  & $54.44$ & $55.62$ &$53.10$ \\
$\Delta c_{4,1}$  & $10.94\%$ & $13.33\%$ &  $8.19\%$  \\
\hline
\end{tabular}
\caption{Predicted fourth-order coefficient $c_{4,1}$ obtained from  the FOPT expansion of $\delta^{(0)}$ using the Levin–T transformation, $\mathcal{L}_{\mathrm{T},k}^{(n)}(\beta)$. The results from the simplified form, $\tilde{\mathcal{L}}_{\mathrm{T},2}$, are already 
reported in Table \ref{tab:coeff4_LevinU_delta0}, and   are not repeated in the current table. Since $\tilde{\mathcal{L}}_{\mathrm{D},1}$ and $\mathcal{L}_{\mathrm{D},1}^{(0)}$ yield the same prediction as $\mathcal{L}_{\mathrm{T},1}^{(1)}$, they are not presented separately in the Levin–D tables. The percentage deviation from the exact value is also shown.}
\label{tab:coeff4_LevinT_delta0}
\end{table}

\begin{table}[H]
\centering
\renewcommand{\arraystretch}{1.4}
\setlength{\tabcolsep}{10pt}
\begin{tabular}{c|ccc}
\hline
Coefficient     & $\mathcal{L}_{\mathrm{T},2}^{(0)}(1)$ & $\mathcal{L}_{\mathrm{T},1}^{(1)}/\tilde{\mathcal{L}}_{\mathrm{D},1}/\mathcal{L}_{\mathrm{D},1}^{(0)}$& $\mathcal{L}_{\mathrm{T},2}^{(0)}(5)$ \\
\hline\hline
$c_{5,1}$    & $279.01$ &$276.15$ & $277.40$  \\
$c_{6,1}$    & $3752.62$ & $3864.54$ & $3636.92$ \\
$c_{7,1}$     & $2.00 \e4$ &$1.95\e4$ &$1.99\e4$  \\
$c_{8,1}$     & $4.12\e5$ &$4.29\e5$ &$3.96\e5$  \\
$c_{9,1}$     & $1.45\e6$ &$1.29\e6$ &$1.49\e6$ \\
$c_{10,1}$    & $7.63\e7$ & $8.08\e7$&$7.27\e7$ \\
$c_{11,1}$   & $-1.73\e8$ & $-2.45\e8$ &$-1.39\e8$ \\
$c_{12,1}$   & $2.47\e{10}$ &$2.67\e{10}$ &$2.33\e{10}$  \\
Levin-Sum   & $0.2122$  & $0.2129$ & $0.2111$ \\
\hline
\end{tabular}
\caption{Predicted higher-order coefficients $c_{5,1}$–$c_{12,1}$ obtained from the FOPT expansion of $\delta^{(0)}$, estimated using the Levin-T sequence transformation, $\mathcal{L}_{\mathrm{T},k}^{(n)}(\beta)$. The results from the  simplified form, $\tilde{\mathcal{L}}_{\mathrm{T},2}$, using three known coefficients as input, are given in Table  \ref{tab:coeff_LevinU_1_delta0}. Since $\tilde{\mathcal{L}}_{\mathrm{D},1}$ and $\mathcal{L}_{\mathrm{D},1}^{(0)}$ yield the same prediction as $\mathcal{L}_{\mathrm{T},1}^{(1)}$, they are not presented separately in the Levin–D tables.}
\label{tab:coeff_LevinT_1_delta0}
\end{table}

In Table~\ref{tab:coeffs_LevinU_2_delta0}, we present the estimates of the higher-order coefficients $c_{5,1}$–$c_{12,1}$ obtained by applying the Levin–U transformation to the FOPT expansion of $\delta^{(0)}$ in Eq.~\eqref{Eq:delta_FOPT}, using the four exactly known coefficients $c_{1,1}$–$c_{4,1}$ as input. The $n=1,k=2$ Levin–U transform is independent of the value  $\beta$. As can be seen from Table~\ref{tab:coeff4_LevinU_delta0}, the performance of the Levin–U transformation deteriorates when the maximal value of $n$ is used, namely, $n=1$ when only three known coefficients are employed as input. This trend persists even when four known coefficients are used as input and the maximal value of $n$ is now $n=2$. Since these cases display poor convergence properties, the corresponding estimates for $c_{5,1}$–$c_{12,1}$ are not included in our final quoted values for the higher-order coefficients.

The corresponding results for the Levin–T and Levin–D transformations are shown in Tables~\ref{tab:coeff_LevinT_2_delta0} and~\ref{tab:coeffs_LevinD_2_delta0}, respectively. For the $n=2,k=1$ case, the Levin–T transform doesn't depend on choice of $\beta$. Likewise, the $n=k=1$ Levin–D transform is independent of any value of $\beta$  and coincides with the $n=2,k=1$ Levin–T result; consequently, separate values for this case are not listed to avoid redundancy. It is evident from these results that the predicted coefficients exhibit very weak dependence on the parameter $\beta$, underscoring the stability and reliability of the Levin-type transformations when applied to the FOPT series of $\delta^{(0)}$.

\begin{table}[h]
\centering
\renewcommand{\arraystretch}{1.4}
\setlength{\tabcolsep}{10pt}
\begin{tabular}{c|ccccc}
\hline
Coefficient & $\mathcal{L}_{\mathrm{U},3}^{(0)}(1)$& $\mathcal{L}_{\mathrm{U},2}^{(1)}$& $\mathcal{L}_{\mathrm{U},1}^{(2)}(1)$ & $\mathcal{L}_{\mathrm{U},3}^{(0)}(5)$   & $\mathcal{L}_{\mathrm{U},1}^{(2)}(5)$\\
\hline\hline
$c_{5,1}$  & $269.53$ & $273.157$&$508.88$ & $264.10$& $392.21$ \\
$c_{6,1}$  & $3287.68$ &$3276.88$ &$1830.47$ & $3287.51$& $2516.2$ \\
$c_{7,1}$  & $1.92\e4$ & $1.97\e4$&$6.00\e4$ & $1.85\e4$&   $4.01\e4$\\
$c_{8,1}$  & $3.48\e5$ &$3.45\e5$ &$-3.46\e4$ & $3.50\e5$& $1.47\e5$ \\
$c_{9,1}$  & $1.53\e6$ & $1.63\e6$&$1.10\e7$ &  $1.40\e6$& $6.35\e6$ \\
$c_{10,1}$ & $6.21\e7$ & $6.09\e7$&$-7.54\e7$ & $6.31\e7$&  $-9.46\e6$ \\
$c_{11,1}$ & $-5.84\e7$ & $-2.23\e7$&$3.38\e9$ & $-9.97\e7$& $1.71\e9$ \\
$c_{12,1}$ & $1.93\e{10}$ &$1.86\e{10}$ &$-5.16\e{9}$ & $2.00\e{10}$& $-1.74\e{10}$ \\
Levin Sum &  $0.2073$  & $0.2075$ & $0.2226$  & $0.2068$& $0.2143$\\
\hline
\end{tabular}
\caption{Predicted higher-order coefficients $c_{5,1}$–$c_{12,1}$ obtained from the FOPT expansion of $\delta^{(0)}$, estimated using the Levin-U sequence transformation, $\mathcal{L}_{\mathrm{U},k}^{(n)}(\beta)$ using four known coefficients as input.}
\label{tab:coeffs_LevinU_2_delta0}
\end{table}

\begin{table}[H]
\centering
\renewcommand{\arraystretch}{1.4}
\setlength{\tabcolsep}{6pt}
\begin{tabular}{c|cccccc}
\hline
Coefficient & $\tilde{\mathcal{L}}_{\mathrm{T},3}$  & $\mathcal{L}_{\mathrm{T},3}^{(0)}(1)$ & $\mathcal{L}_{\mathrm{T},2}^{(1)}(1)$&$\mathcal{L}_{\mathrm{T},1}^{(2)}/\mathcal{L}_{\mathrm{D},1}^{(1)}$ & $\mathcal{L}_{\mathrm{T},3}^{(0)}(5)$ & $\mathcal{L}_{\mathrm{T},2}^{(1)}(5)$ \\
\hline\hline
$c_{5,1}$   & $258.66$ & $284.08$ & $288.93$& $304.71$ & $270.45$ & $281.05$  \\
$c_{6,1}$   & $3267.68$ & $3281.57$ & $3262.00$&$3171.08$ & $3301.05$ & $3278.94$ \\
$c_{7,1}$   &$ 1.78 \e 4$ & $2.12\e4$ & $2.19\e4$& $2.44\e4$ & $1.93\e4$ & $2.08\e4$\\
$c_{8,1}$   & $3.47 \e{5} $ & $3.43\e5$ & $3.38\e5$&$3.15\e5$ & $3.50\e5$ & $3.43\e5$\\
$c_{9,1}$   & $1.27\e6$ & $1.91\e6$ & $2.06\e6$ & $2.63\e6$ &$1.55\e6$ & $1.84\e6$\\
$c_{10,1}$  & $6.32\e7$ & $5.92\e7$ & $5.71\e7$ & $4.90\e7$&$6.26\e7$ & $5.95\e7$\\
$c_{11,1}$  & $-1.34\e8$ & $6.23\e7$ & $1.16\e8$ & $3.22\e8$&$-5.64\e7$ & $4.17\e7$\\
$c_{12,1}$  & $2.02\e{10}$ & $1.74\e{10}$ & $1.64\e{10}$ & $1.22\e{10}$ & $1.95\e{10}$ & $1.77\e{10}$ \\
Levin Sum & $0.2059$     &  $0.2088$   & $0.2092$& $0.2100$  &  $0.2076$    & $0.2085$ \\
\hline
\end{tabular}
\caption{Predicted higher-order coefficients $c_{5,1}$–$c_{12,1}$ obtained from the FOPT expansion of $\delta^{(0)}$, estimated using the Levin-T sequence transformation, $\mathcal{L}_{\mathrm{T},k}^{(n)}(\beta)$, and its simplified form, $\tilde{\mathcal{L}}_{\mathrm{T},k}$,  using four known coefficients as input. Since  $\mathcal{L}_{\mathrm{D},1}^{(1)}$ yield the same prediction as $\mathcal{L}_{\mathrm{T},1}^{(2)}$, they are not presented separately in the Levin–D tables.}
\label{tab:coeff_LevinT_2_delta0}
\end{table}

\begin{table}[H]
\centering
\renewcommand{\arraystretch}{1.4}
\setlength{\tabcolsep}{10pt}
\begin{tabular}{c|ccc}
\hline
Coefficient & $\tilde{\mathcal{L}}_{\mathrm{D},2}$  & $\mathcal{L}_{\mathrm{D},2}^{(0)}(1)$  & $\mathcal{L}_{\mathrm{D},2}^{(0)}(5)$ \\
\hline\hline
$c_{5,1}$  & $273.16$ & $294.19$  & $282.17$ \\
$c_{6,1}$  & $3276.88$ & $3240.14$  & $3277.68$ \\
$c_{7,1}$  & $1.97\e4$ & $2.26\e4$  & $2.10\e4$ \\
$c_{8,1}$  & $3.45\e5$ & $3.32\e5$  & $3.43\e5$ \\
$c_{9,1}$  & $1.64\e6$ & $2.23\e6$   & $1.87\e6$ \\
$c_{10,1}$ & $6.09\e7$ & $5.50\e7$  & $5.92\e7$ \\
$c_{11,1}$ & $-2.23\e7$ & $1.74\e8$  & $5.15\e7$ \\
$c_{12,1}$ & $1.86\e{10}$ & $1.53\e{10}$  & $1.76\e{10}$ \\
Levin Sum & $0.2075$& $0.2095$ &$0.2086$  \\
\hline
\end{tabular}
\caption{Predicted higher-order coefficients $c_{5,1}$–$c_{12,1}$ obtained from the FOPT expansion of $\delta^{(0)}$, estimated using the Levin-D sequence transformation $\mathcal{L}_{\mathrm{D},k}^{(n)}(\beta)$ and it's simplified form $\tilde{\mathcal{L}}_{\mathrm{D},k}$, using four known coefficients as input.}
\label{tab:coeffs_LevinD_2_delta0}
\end{table}

\section{Final predictions of the higher order coefficients $c_{5,1}$–$c_{12,1}$ }
\label{higher_orders}
Our final predictions for the higher-order coefficients $c_{5,1}$-$c_{12,1}$
are obtained by taking the mean of a selected set of estimates listed in
Tables~\ref{tab:coeff_LevinU_1_delta0},
\ref{tab:coeff_LevinT_1_delta0}, and
\ref{tab:coeffs_LevinU_2_delta0}-\ref{tab:coeffs_LevinD_2_delta0}.
These tables contain results derived from the application of Levin-type
sequence transformations to the FOPT expansion of $\delta^{(0)}$, using both
three known input coefficients and four known input coefficients.

For the case where three known input coefficients are employed, we first
consider the results presented in Table~\ref{tab:coeff_LevinU_1_delta0}. From
this table, we exclude the estimates obtained using the
$\mathcal{L}_{\mathrm{U},1}^{(1)}(1)$ and
$\mathcal{L}_{\mathrm{U},1}^{(1)}(5)$ transformations, which exhibit
substantially larger deviations in the prediction of the known fourth-order
coefficient $c_{4,1}$. All remaining entries from
Table~\ref{tab:coeff_LevinU_1_delta0} are included in the averaging procedure.
We then include all entries from Table~\ref{tab:coeff_LevinT_1_delta0}, which
provides the corresponding results obtained from the Levin-$\rm T$ transformations.

When four known input coefficients are used, we again omit the
$\mathcal{L}_{\mathrm{U},1}^{(2)}(1)$ and
$\mathcal{L}_{\mathrm{U},1}^{(2)}(5)$ columns from
Table~\ref{tab:coeffs_LevinU_2_delta0}, consistent with the selection criteria
adopted earlier. For the corresponding cases of the Levin-T and Levin-D transformations,
shown in Tables~\ref{tab:coeff_LevinT_2_delta0} and
\ref{tab:coeffs_LevinD_2_delta0}, respectively, all columns are included in the
final averaging procedure.

This selection ensures that the final estimates for $c_{5,1}$-$c_{12,1}$ are
based on a set of transformations that yield mutually consistent predictions
for the known coefficient $c_{4,1}$, while excluding only those cases that lead
to substantially larger deviations, thereby providing a stable and reliable
determination of the higher-order coefficients.

The error associated with each estimated coefficient is defined as the maximum spread of the predicted values, obtained from the different Levin-type transformations (U, T, D), providing a conservative estimate of the error associated with the choice of transformation.  A comparison of the final predicted coefficients with the  results from Refs.~\cite{Boito:2018rwt,BeJa} is demonstrated in Table \ref{tab:final_coeff}. Our predictions are in  good agreement with them, 
demonstrating that the estimates obtained from the Levin transformations of  the FOPT expansion of $\delta^{(0)}$ are consistent with previous studies.

\begin{table}[H]
\centering
\renewcommand{\arraystretch}{1.4}
\setlength{\tabcolsep}{10pt}
\begin{tabular}{c|ccc}
\hline
Coeff. & This work & Ref.~\cite{Boito:2018rwt} & Ref.~\cite{BeJa} \\
\hline
$c_{5,1}$  & $278^{+27}_{-19}$  & $277\pm 51$ & $283$ \\
$c_{6,1}$  & $3375^{+489}_{-204}$ & $3460\pm 690$ & $3275$ \\
$c_{7,1}$  & $(2.03^{+0.41}_{-0.25})\e4$ &  $(2.02\pm 0.72)\e4$ & $1.88 \e 4$ \\
$c_{8,1}$  & $(3.6^{+0.7}_{-0.4})\e5$ &  $(3.7\pm 1.1)\e5$ & $3.88 \e5$ \\
$c_{9,1}$  & $(1.7^{+0.9}_{-0.4})\e6$ & $(1.6\pm 1.4)\e6$ & $9.19 \e5$ \\
$c_{10,1}$ & $(6.3^{+1.8}_{-1.4})\e7$ & $(6.6\pm 3.2)\e7$ & $8.37 \e7$ \\
$c_{11,1}$ & $(-2.0^{+34}_{-22})\e7$ & $(-5\pm 57)\e7$ & $-5.19 \e8 $ \\
$c_{12,1}$ & $(1.9^{+0.7}_{-0.7})\e{10}$ & $(2.1\pm 1.5)\e{10}$ & $3.38\e{10} $ \\
\hline
\end{tabular}
\caption{Comparison of predicted coefficients $c_{n,1}$ ($n=5$–$12$) with results from Refs.~~\cite{Boito:2018rwt} and ~\cite{BeJa}.}
\label{tab:final_coeff}
\end{table}

\section{An estimate  of $\delta^{(0)}$}   
\label{delta0}
In this section, we provide our final estimate of the correction $\delta^{(0)}$.  For this purpose, we use the value of the $\delta^{(0)}$ predicted by the Levin sum of the FOPT series of $\delta^{(0)}$  given in Eq. \eqref{Eq:delta_FOPT}.  The final result using $\alpha_s(M_\tau)=0.31959 \pm 0.00445$ is,
\begin{equation}
   \delta^{(0) }_{\text{Levin-FOPT}}=0.2089 \pm 0.0040\pm 0.0060_{\alpha_s}.
   \label{delta0_best}
\end{equation}
The first uncertainty is due to spread of Levin-sums of $\delta^{(0)}$ obtained from various Levin-transformation and the second arises from the uncertainty in $\alpha_s$.

\begin{figure}[H]
	\centering
	 \includegraphics[width=0.5\linewidth]{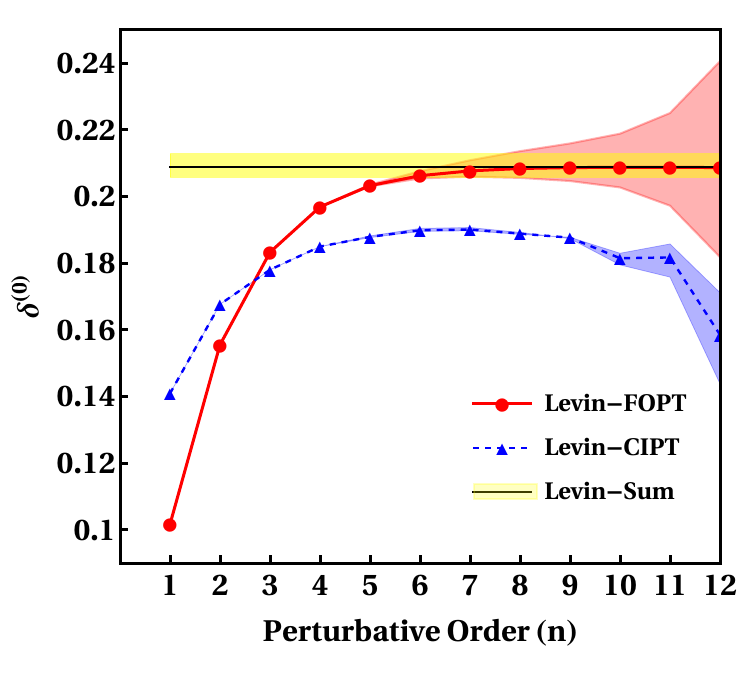}
 \caption{
 Final prediction of $\delta^{(0)}$ in QCD  using the higher-order coefficients listed in Table~\ref{tab:final_coeff}. The shaded regions in the perturbative expansions represent the uncertainties from the coefficients. The solid black curves corresponds to the mean of the Levin-summed values of $\delta^{(0)}$. The shaded yellow bands denote the spread of the Levin sums obtained from different Levin-type transformations (U, T, D). The input values $\alpha_s=0.31959$ is used. }
  \label{fig_4}
	\end{figure}

In addition, our final prediction of $\delta^{(0)}$ in QCD is shown in Fig.~\ref{fig_4}, where we use the coefficients $c_{5,1}$–$c_{12,1}$ from Table~\ref{tab:final_coeff}. It is evident that the FOPT series begins to align closely with the Levin-summed result from the sixth order onward, indicating improved convergence at higher orders.

While the our  best  $\overline{\text{MS}}$ estimate of $\delta^{(0)}$ is the value reported in Eq.~\ref{delta0_best}, a direct comparison with earlier studies in the literature necessitates adopting the same input value of $\alpha_s(M_\tau)$ used in those analyses. Accordingly, Table~\ref{tab:delta_0_lit} also includes an alternative Levin-transform prediction obtained using the identical $\alpha_s(M_\tau)$ input as in the previous works. The level of agreement among our results and previous determinations is remarkable. This consistency highlights the reliability and robustness of Levin-type sequence transformations when applied to perturbative QCD.

\begin{table}[h!]
\centering
\renewcommand{\arraystretch}{1.6}
\small
\begin{adjustbox}{max width=\textwidth}
\begin{tabular}{p{1cm}|p{2.7cm}|p{2.7cm}|p{2.7cm}|p{2.7cm}|p{2.7cm}|p{2.7cm}}
\hline
 & This work 
 & Ref.~\cite{Boito:2018rwt} 
 & Ref.~\cite{BeJa} 
 & Ref.~\cite{Boito:2016pwf} 
 & Ref.~\cite{Wu:2018cmb} 
 & Ref.~\cite{Caprini:2018agy} \\
\hline\hline
$\delta^{(0)}$  
& $0.2043 \pm 0.0037 \pm 0.0130_{\alpha_s}$  
& $0.2050 \pm 0.0067 \pm 0.0130_{\alpha_s}$  
& $0.2047 \pm 0.0029 \pm 0.0130_{\alpha_s}$  
& $0.2047 \pm 0.0034 \pm 0.0133_{\alpha_s}$  
& $0.2035 \pm 0.0030 \pm 0.0123_{\alpha_s}$  
& $0.2018 \pm 0.0211 \pm 0.0123_{\alpha_s}$ \\
\hline
\end{tabular}
\end{adjustbox}
\caption{Comparison of the estimates of $\delta^{(0)}$ obtained in this work with results reported in the literature. All values are evaluated using a common input $\alpha_s(M_\tau)=0.316 \pm 0.010$.}
\label{tab:delta_0_lit}
\end{table}

\section{Summary and conclusion}

In this work, we have, for the first time, applied the Levin-type sequence transformations  to the estimation of higher-order perturbative QCD corrections to hadronic $\tau$ decays. The hadronic $\tau$ decays provide an exceptionally clean and well-understood low-energy probe of QCD, where both perturbative and non-perturbative effects can be quantitatively analyzed. On the other hand, the Levin-type sequence transformations represent a powerful class of nonlinear convergence accelerators that incorporate explicit information about the asymptotic behavior of the remainder terms of a series. By constructing weighted combinations of successive partial sums, they effectively suppress truncation errors and enhance the convergence properties of slowly convergent or even divergent perturbative series.

%We first apply the Levin-type transformations to the inverted series expansion of $\alpha_s$ defined in Eq.~\eqref{Eq:aFOPT}. This series exhibits strong divergence and large-order oscillations, rendering direct summation unstable and physically unreliable. The application of Levin-type transformations to this series leads to a remarkable acceleration of convergence, producing a stable and precise prediction for $\alpha_s(M_\tau^2)$. In addition, the Levin-summed inverted series provides a means to estimate higher-order QCD corrections to hadronic $\tau$ decays beyond the known fixed-order results.

% We apply the Levin-transformations to the perturbative expansion of the quantity $\delta^{(0)}$, which governs the dominant perturbative contribution to the hadronic $\tau$ width. This procedure enables the extraction of a range of $\alpha_s(M_\tau^2)$ values that reproduce the experimentally measured $\delta^{(0)}$. The extracted range of $\alpha_s(M_\tau^2)$ is found to be in excellent agreement with the value obtained independently from the Levin-summed inverted series of $\alpha_s$, lending strong support to the internal consistency of the method. The resulting determination constitutes our final prediction for the strong coupling constant at the $\tau$ mass scale.

We apply the Levin-transformations to the perturbative expansion of the quantity $\delta^{(0)}$, which governs the dominant perturbative contribution to the hadronic $\tau$ width. The application of Levin-type transformations to the perturbative series of $\delta^{(0)}$ allows us to estimate the higher-order coefficients $c_{5,1}$–$c_{12,1}$. Our predictions for $c_{5,1}$–$c_{12,1}$ show excellent consistency with the most reliable estimates available in the literature, confirming the robustness and predictive power of the Levin-type transformations in the context of QCD perturbation theory.

The present analysis demonstrates that Levin-type sequence transformations constitute a highly effective mathematical tool for improving the convergence properties of perturbative QCD series relevant to hadronic $\tau$ decays. Their successful application to  the  perturbative expansions provides a reliable predictions for higher-order coefficients. These results highlight the capability of nonlinear Levin-sequence transformations to systematically extend the predictive reach of QCD beyond the currently known orders.

Furthermore, Levin-type sequence transformations open a promising new direction for the perturbative treatment within FESR analyses used to extract $\alpha_s$ from inclusive hadronic $\tau$ decays. Their demonstrated ability to stabilize and reorganize divergent series suggests that they could meaningfully advance the precision frontier in strong-coupling determinations. As a natural continuation of this work, we plan to develop a dedicated, fully fledged FESR framework, in near future,  built on these transformations, with the broader goal of establishing them as a robust tool for precision QCD phenomenology.

From a broader perspective, the present work opens several promising directions. First, the method can be readily generalized to other observables where perturbative expansions are known to be slowly convergent or factorially divergent, such as the Bjorken sum rule, or the Higgs hadronic decay width. Second, the interplay between Levin-type transformations and Borel summation techniques offers an intriguing avenue for bridging sequence transformations with renormalon-based analyses, potentially allowing a quantitative comparison with resummed perturbation theory or analytic QCD models. 

In summary, this study establishes the Levin-type transformations as a powerful and versatile framework for addressing divergence and truncation problems in perturbative quantum field theory. Their ability to accelerate convergence, predict higher-order terms, and yield physically consistent results suggests that they may serve as an indispensable bridge between perturbative QCD and its resummed or nonperturbative formulations.

\label{sum}

\section*{Acknowledgments}
 This work is supported by the  Council of Science and Technology,  Govt. of Uttar Pradesh,  India, through the project ``   A new paradigm for flavor problem "  no.   CST/D-1301,  and Anusandhan National Research Foundation (Science and Engineering Research Board) , Department of Science and Technology, Government of India through the project ``Higgs Physics within and beyond the Standard Model" no. CRG/2022/003237.  

\begin{appendix}

\end{appendix}

\end{document}